\documentclass[preprint,authoryear,12pt]{elsarticle}

\usepackage{epsfig}

\usepackage{amssymb}

\journal{Advances in Space Research}

\makeatletter
\def\@author#1{\g@addto@macro\elsauthors{\normalsize%
    \def\baselinestretch{1}%
    \upshape\authorsep#1\unskip\textsuperscript{%
      \ifx\@fnmark\@empty\else\unskip\sep\@fnmark\let\sep=,\fi
      \ifx\@corref\@empty\else\unskip\sep\@corref\let\sep=,\fi
      }%
    \def\authorsep{\unskip,\space}%
    \global\let\@fnmark\@empty
    \global\let\@corref\@empty  
    \global\let\sep\@empty}%
    \@eadauthor={#1}
}
\makeatother

\newcounter{nbdrafts}
\makeatletter
\newcommand{\checknbdrafts}{
\ifnum \thenbdrafts > 0
\@latex@warning@no@line{*****************************************************************}
\@latex@warning@no@line{*WARNING* The document contains \thenbdrafts \space draft note(s)}
\@latex@warning@no@line{*****************************************************************}
\fi}

\makeatother

\usepackage{hyperref}

\usepackage[multiple]{footmisc}

\usepackage{color, colortbl}
\definecolor{gray}{rgb}{0.8,0.8,0.8}

\usepackage{csvsimple} 

\usepackage{array}

\usepackage{wrapfig}

\usepackage{subcaption}

\usepackage{caption}

\biboptions{authoryear}

\begin{document}

\begin{frontmatter}

\title{Geometric calibration of \textbf{C}olour \textbf{a}nd \textbf{S}tereo \textbf{S}urface \textbf{I}maging \textbf{S}ystem of ESA's \textbf{T}race \textbf{G}as \textbf{O}rbiter}

\author[epfl]{Stepan Tulyakov\corref{cor}}
\cortext[cor]{Corresponding author}
\ead{stepan.tulyakov@epfl.ch}

\author[epfl]{Anton Ivanov}
\ead{anton.ivanov@epfl.ch}

\author[unibern]{Nicolas Thomas}
\ead{nicolas.thomas@space.unibe.ch}

\author[unibern]{Victoria Roloff}
\ead{victoria.roloff@space.unibe.ch}

\author[unibern]{Antoine Pommerol}
\ead{antoine.pommerol@space.unibe.ch}

\author[padova]{Gabriele Cremonese}
\ead{gabriele.cremonese@oapd.inaf.it}

\author[ruag]{Thomas Weigel}
\ead{thomas.weigel@ruag.com}

\author[idiap]{Francois Fleuret}
\ead{francois.fleuret@idiap.ch}

\address[epfl]{\' Ecole Polytechnique Federale de Lausanne, EPFL-ENT-ESC, Station 13, 1015 Lausanne, Switzerland}
\address[unibern]{Physics Institute, Space Research and Planetary Sciences, University of Bern, Sidlerstrasse 5,
3012 Bern, Switzerland}
\address[padova]{INAF, Osservatorio Astronomico di Padova, Vicolo Osservatorio 5, 35122 Padova, Italy}
\address[ruag]{RUAG Space, RUAG Schweiz AG Schaffhauserstrasse 580, 8052 Zürich, Switzerland}
\address[idiap]{Idiap Research Institute, Rue Marconi 19, 1920 Martigny, Switzerland}

\begin{abstract}
There are many geometric calibration methods for ``standard'' cameras. These methods, however, cannot be used for the calibration of telescopes with large focal lengths and complex off-axis optics. Moreover, specialized calibration methods for the telescopes are scarce in literature. We describe the calibration method that we developed for the Colour and Stereo Surface Imaging System~(CaSSIS) telescope, on board of the ExoMars~Trace Gas Orbiter~(TGO). Although our method is described in the context of CaSSIS, with camera-specific experiments, it is general and can be applied to other telescopes. We further encourage re-use of the proposed method by making our calibration code and data available on-line.
\end{abstract}


\end{frontmatter}

\parindent=0.5 cm

\section{Introduction}
\label{sec:intro}

\begin{figure}
\centering
\begin{minipage}{.45\textwidth}
  \centering
  \includegraphics[trim={0cm 0cm 0cm 0cm},clip,width=0.99\textwidth]{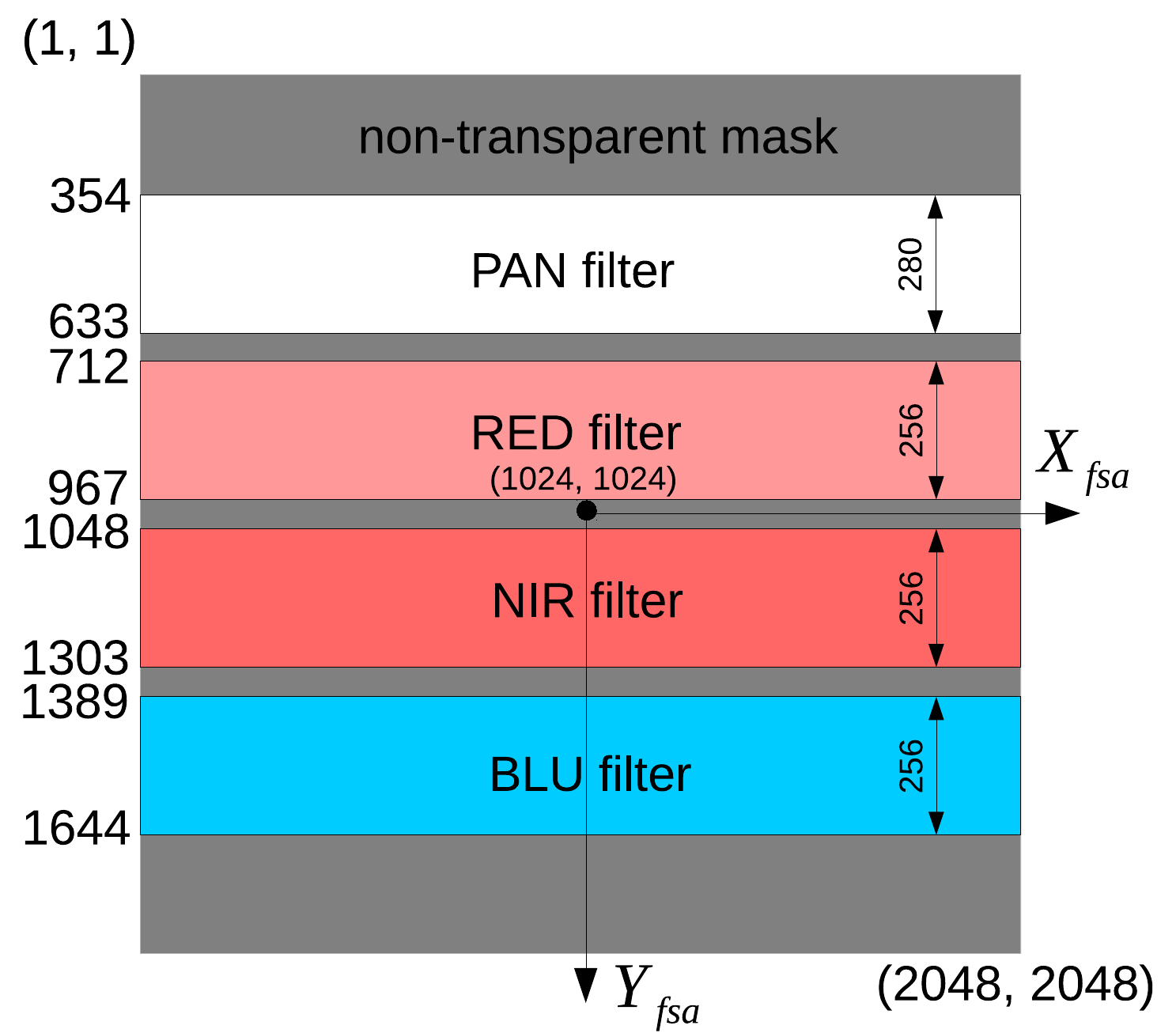}
\end{minipage}%
\begin{minipage}[b]{.55\textwidth}
\centering \resizebox{1\columnwidth}{!}{%
  \scriptsize
  \begin{tabular}{ll}
  \hline
  Quantity & Value  \\
  \hline
  Optic  & 3-mirror plus fold mirror off-axis \\
  Detector &  Raytheon Osprey 2k CMOS hybrid \\
  Filters & 675 nm, 485 nm, 840 nm, 985 nm \\
  Focal length & 880 mm  \\
  F\#           & 6.52    \\
  Pixel size   & 10 um x 10 um   \\
  Detector area & 2048 x 2048 (2048 x 1350 is used)\\
  FOV & 1.33 deg x 0.88 deg\\
  Angle w.r.t nadir & 10.0 +/- 0.2 deg \\ 
  \hline
  \end{tabular}
  }
  \end{minipage}
  \caption{CaSSIS camera specification. The CaSSIS telescope is three-mirror anastigmat system (off-axis) with a fold mirror. The CaSSIS Filter Strip Assembly~(FSA) comprises a Raytheon Osprey 2048$\times$2048 hybrid CMOS detector with 4 colour filters mounted on it following the push-frame technique. Small dark bands between the filters reduce spectral cross-talk. The detector can acquire an un-smeared image along  ground-track. The along-track dimension of the image is then built up and put together on ground.}%
\label{fig:cassis_parms}
\end{figure}

On March 15, 2016 Trace Gas Orbiter~(TGO) was launched to Mars, as part of the the European Space Agency's~(ESA's) ExoMars project. Its aim is to find trace gases, which may be evidence of geological or biological activity on Mars. The Colour and Stereo Surface Imaging System~(CaSSIS) is TGO's imaging system that provides visual context for sites identified as potential sources of  trace gases. A brief specification of CaSSIS is provided in Tab.~\ref{fig:cassis_parms}.

CaSSIS~(\citealt{Thomas2014, Thomas2017}) is a multi-spectral push-frame camera with 4 rectangular color filters covering its sensor (Fig.~\ref{fig:cassis_parms}). As the spacecraft is moving along the orbit, each part of a targeted area becomes visible, sequentially, in each filter. By acquiring and mosaicking multiple images (``framelets''), CaSSIS is able to reconstruct a large 4-colours image of the targeted area.

CaSSIS is also a stereo camera. It is capable of acquiring two images of a target area from two distinct points on the same orbit. While approaching the target area it acquires the first image, then it gets mechanically rotated and acquires the second image, while departing from the target area. By computing parallax from these two images, one can reconstruct a Digital Elevation Model~(DEM) of the target area.

To prepare scientific products, such as color images and DEMs from raw CaSSIS images, one needs geometric camera parameters, such as its focal length and a optical distortion model. While their nominal values are known from technical specification, their actual values might deviate from the nominal ones, due to imprecise manufacturing, mounting, or various incidents during the spacecraft cruise and operation. Therefore, their actual values have to be measured in the controlled environment of the clean room and validated during the commissioning phase in flight. This is the main goal of geometric calibration. Note that photometric calibration of CaSSIS is described in \citealt{Roloff2017}.

There are many geometric calibration methods~(\citealt{Hartley2003,Zhang1999, Heikkila97, Tsai1987}) and tools\footnote{MATLAB camera calibration tool, \url{https://ch.mathworks.com/help/vision/ug/single-camera-calibrator-app.html}, accessed 2017-05-23}\footnote{OpenCV camera calibration tool, \url{http://docs.opencv.org/2.4/doc/tutorials/calib3d/camera_calibration/camera_calibration.html}, accessed 2017-05-23}\footnote{Caltech camera calibration tool, \url{http://www.vision.caltech.edu/bouguetj/calib_doc/}, accessed 2017-05-23} for ``standard'' cameras. However, these off-the-shelf tools cannot be used for the calibration of telescopes such as CaSSIS, for two reasons. Firstly, most of these tools require images of calibration targets, such as a checkerboard chart. For telescopes with a large focal length, however, such targets must be very large ($\approx km^2$) and should be placed very far away from the telescope ($\approx 10\ km $), which is impractical. Secondly, telescopes often have off-axis optical designs with complex optical distortion, that cannot be handled by off-the-shelf tools. Therefore, there is a need for specialized calibration methods, which are unfortunately scarce in the literature.

In this paper we describe the calibration method that we developed for CaSSIS. Although our method is described in the context of CaSSIS, it is general and can be applied to other telescopes. We further encourage re-use of the proposed method by making our calibration code and data available on-line\footnote{\url{https://github.com/eSpaceEPFL/CASSISgeometry.git}}.

We first discuss in~\S~\ref{sec:related_work} the related work, and describe in~\S~\ref{sec:camera_model} the geometric camera model adopted in the paper. In~\S~\ref{sec:distortion_model_selection} we explain the distortion model selection procedure based on lens simulation, in~\S~\ref{sec:on_ground_calibration} we describe the on-ground calibration using images of a dotted calibration target captured through a collimator, and in~\S~\ref{sec:in_flight_calibration} we describe the in-flight calibration using star field images. Finally, in~\S~\ref{sec:color_image_experiment} we show how refined geometric parameters improve the quality of map-projected CaSSIS images.

\section{Related Work}
\label{sec:related_work}

\subsection{Optical Distortion models}
\label{sec:lens_distortion_models}

Off-the-shelf calibration tools typically assume a radial or a Brown-Conrady optical distortion model. The radial model~(\citealt{Hartley2003}) is a simple 5 degrees-of-freedom~(DOF) model, only accounting for radially symmetric distortion. Brown-Conrandy~(\citealt{Brown1966}) is a more complex model with 7 DOF, that in addition to the radially symmetric component, accounts for tangential decentering. These models, however, cannot represent the complex distortion in a camera with off-axis optical elements, such as CaSSIS, as we show in~\S~\ref{sec:distortion_model_selection}.
Complex distortion is better modeled by a bi-cubic~(\citealt{kilpela1981}) or a rational model~(\citealt{claus2005}) with 20 and 17~DOF, respectively. In our work we adopted the rational distortion model (discussed in~\S~\ref{sec:camera_model}).

\subsection{Star field calibration}
\label{sec:starfield_calibration}

For geometric calibration of a camera one needs images of calibration targets - objects with known real-world coordinates. Since angular positions of stars are well known and documented in star catalogs\footnote{VisieR star catalog library, \url{http://vizier.u-strasbg.fr/}, accessed: 2017-05-24}
, such as MASS2 and Tycho2, star fields can serve as perfect calibration targets. Indeed, star field calibration is widely used in star trackers that are an integral part of every spacecraft~(\citealt{Samaan2001, Pal2014, Junfeng2005}). The star field calibration can also be used for calibration of consumer-level cameras~(\citealt{Klaus2004}). Unfortunately, all known star field-based calibration methods assume a simplistic optical distortion model, and therefore, cannot be applied for telescope calibration. Before stars from an image can be used for calibration, they should be identified using a star catalog, which fortunately can be done automatically~(\citealt{lang2009}) with a Astrometry.net library~\footnote{Astrometry.net star recognition tool, \url{http://astrometry.net/}, accessed: 2017-05-24}. 

\section{CaSSIS camera model}
\label{sec:camera_model}

The camera model consists of: (1)~the intrinsic model,
(3)~the rational optical distortion model and (3)~the extrinsic model. In this section we discuss each part of the camera model in detail.

\subsection{Intrinsic Model}

The intrinsic model~(\citealt[p153-158]{Hartley2003}) describes the transformation from 3D camera frame coordinates \( \mathbf{X} = \left\{X, Y, Z \right\} \) to  2D image coordinates \(\mathbf{x} = \left\{x, y\right\}\) as follows:
\begin{equation}
\label{eq:intrinsic_model}
(x, y) = \left( \frac{\mathbf{K}_1^T\mathbf{X}}{\mathbf{K}_3^T\mathbf{X}},
    \frac{\mathbf{K}_2^T\mathbf{X}}{\mathbf{K}_3^T\mathbf{X}} \right), \ \ \
\mathbf{K} =\left[ \begin{array}{ccc}
     f        & 0        & x_0 \\
     0        & f 		 & y_0 \\
     0        & 0        & 1
    \end{array} \right],
\end{equation}
where \(f\) is the focal length of the camera, measured in pixels, and \(x_0\), \(y_0\) are the coordinates of the principal point in the image. In the case of CaSSIS, we assume that \(x_0\) and \(y_0\) correspond to the center of an image. Therefore, the CaSSIS intrinsic model has only 1 DOF.

\subsection{Rational optical distortion model}
\label{sec:rational_model}

The intrinsic camera model is complemented with a optical distortion model, that describes the transformation from the distorted image coordinates \( \mathbf{i}=(i, j) \) to the ideal image coordinates \( \mathbf{x}=(x, y) \). We use the rational distortion model~(\citealt{claus2005}):
\begin{equation}
	\label{eq:6D_lifted_coordinates}
(x, y) = \left( \frac{\mathbf{A}^T_1\mathbf{\chi_6}}	{\mathbf{A}^T_3\mathbf{\chi_6}},                      \frac{\mathbf{A}^T_2\mathbf{\chi_6}}{\mathbf{A}^T_3\mathbf{\chi_6}} \right), \ \ \
    \mathbf{\chi_6} = \left[ \begin{array}{cccccc}
     i^2 & ij & j^2 & i & j & 1 \end{array} \right]^T,
\end{equation}
where \(\mathbf{A}^T_{1...3}\) are rows of a \(3\times6\) rational distortion matrix.

Interestingly, while not invertible, the rational model can represent very precisely the inverse of itself~(\citealt{tang2012}). We use this property and simultaneously estimate two rational models: one for distortion and another for correction.

\subsection{Extrinsic Model}
\label{sec:extrinsic_model}

The extrinsic model~(\citealt[p155-156]{Hartley2003}) describes the transformation from the reference (spacecraft) frame  coordinates \( \mathbf{X_{ref}} = (X_{ref}, Y_{ref}, Z_{ref}) \) to the camera frame coordinates \( \mathbf{X} = (X, Y, Z) \) as follows
\begin{equation}
\label{eq:extrinsics}
\mathbf{X} = \mathbf{R}\mathbf{X_{ref}} + \mathbf{T},
\end{equation}
where \(\mathbf{R}\) is the \(3\times3\) rotation matrix, and $\mathbf{T}$ is the \(3\times1\) translation vector. The rotation matrix \(\mathbf{R}\) is the function of 3 Euler angles \( \mathbf{R} = F(\alpha, \beta, \gamma)\). The translation vector is typically ignored, because the camera is much closer to the reference frame than to the scene. The extrinsic model has 3 DOF in total.

\section{Distortion model selection}
\label{sec:distortion_model_selection}

Before CaSSIS was assembled, we were provided with optical distortion data by the telescope manufacturer (RUAG Space Zurich, Switzerland), shown in Tab.~\ref{tab:lens_simulation}, computed using a ray-tracing simulation. To find out what distortion model better represented CaSSIS optical distortion, we fitted radial, Brown-Conrady, rational and bi-cubic optical distortion models (see~\S~\ref{sec:lens_distortion_models}) to the data, and compared the average Euclidean error of the models using leave-one-out cross-validation.

\begin{figure}[hbtp]
\centering
\begin{subfigure}{.49\textwidth}
  \centering
  \includegraphics[width=1\textwidth]{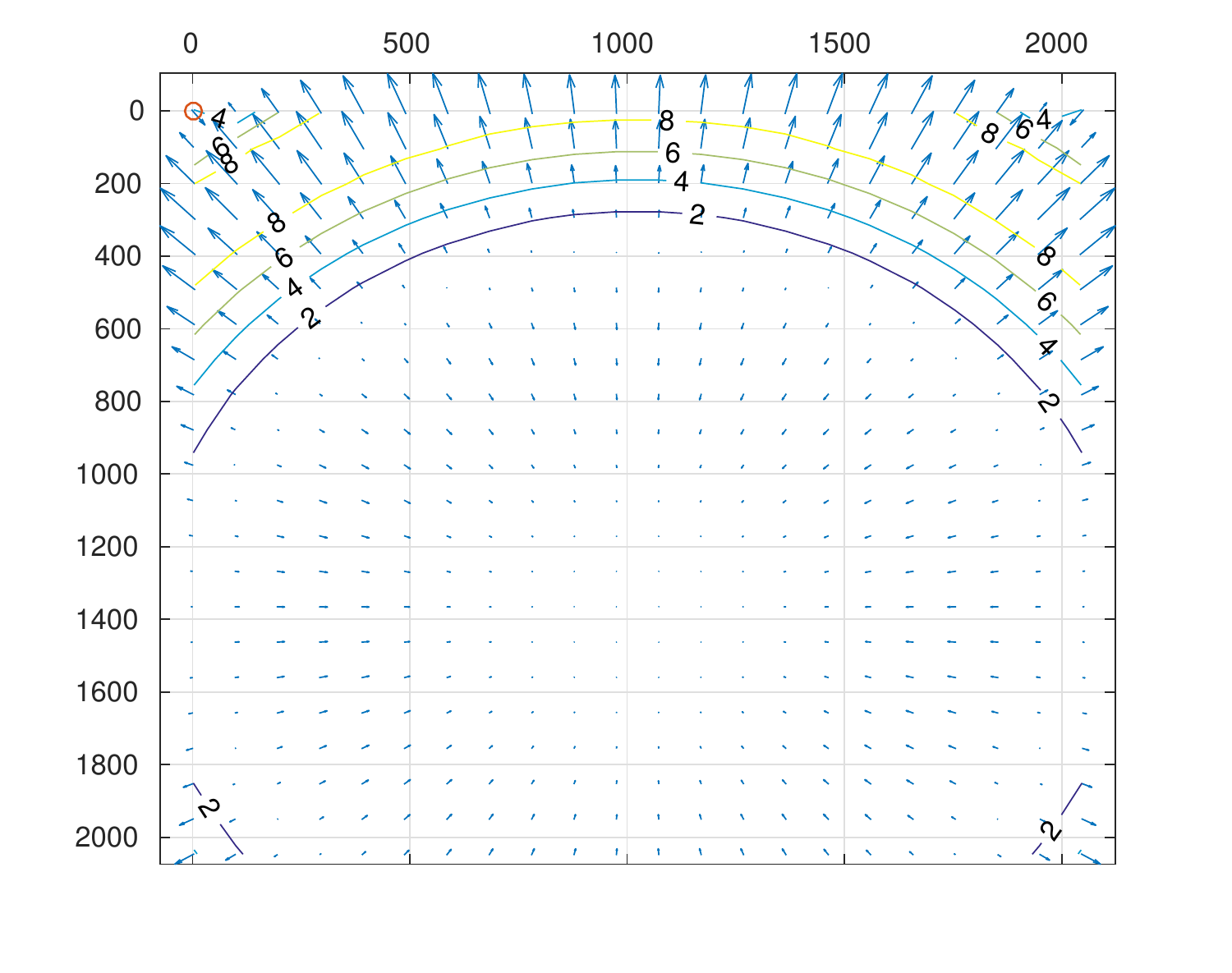}
  \caption{radial (\(\overline{error}\)=3.169 pix)}
  \label{fig:simulation_radial_model}
\end{subfigure}
\begin{subfigure}{.49\textwidth}
  \centering
  \includegraphics[width=1\textwidth]{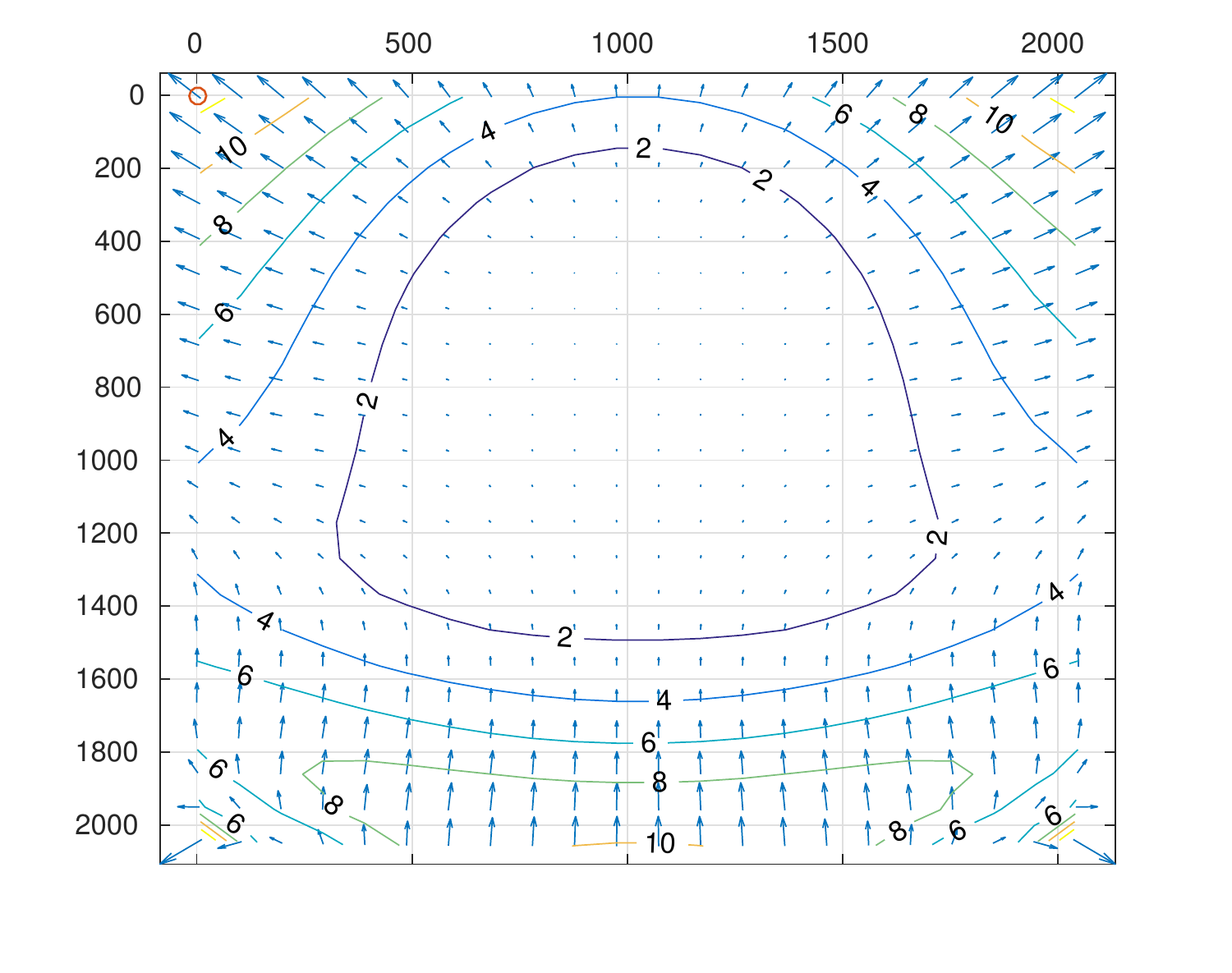}
  \caption{brown-conrandy (\(\overline{error}\)=1.585 pix)}
  \label{fig:simulation_brown_conrandy_model}
\end{subfigure}
\begin{subfigure}{.49\textwidth}
  \centering
  \includegraphics[width=1\textwidth]{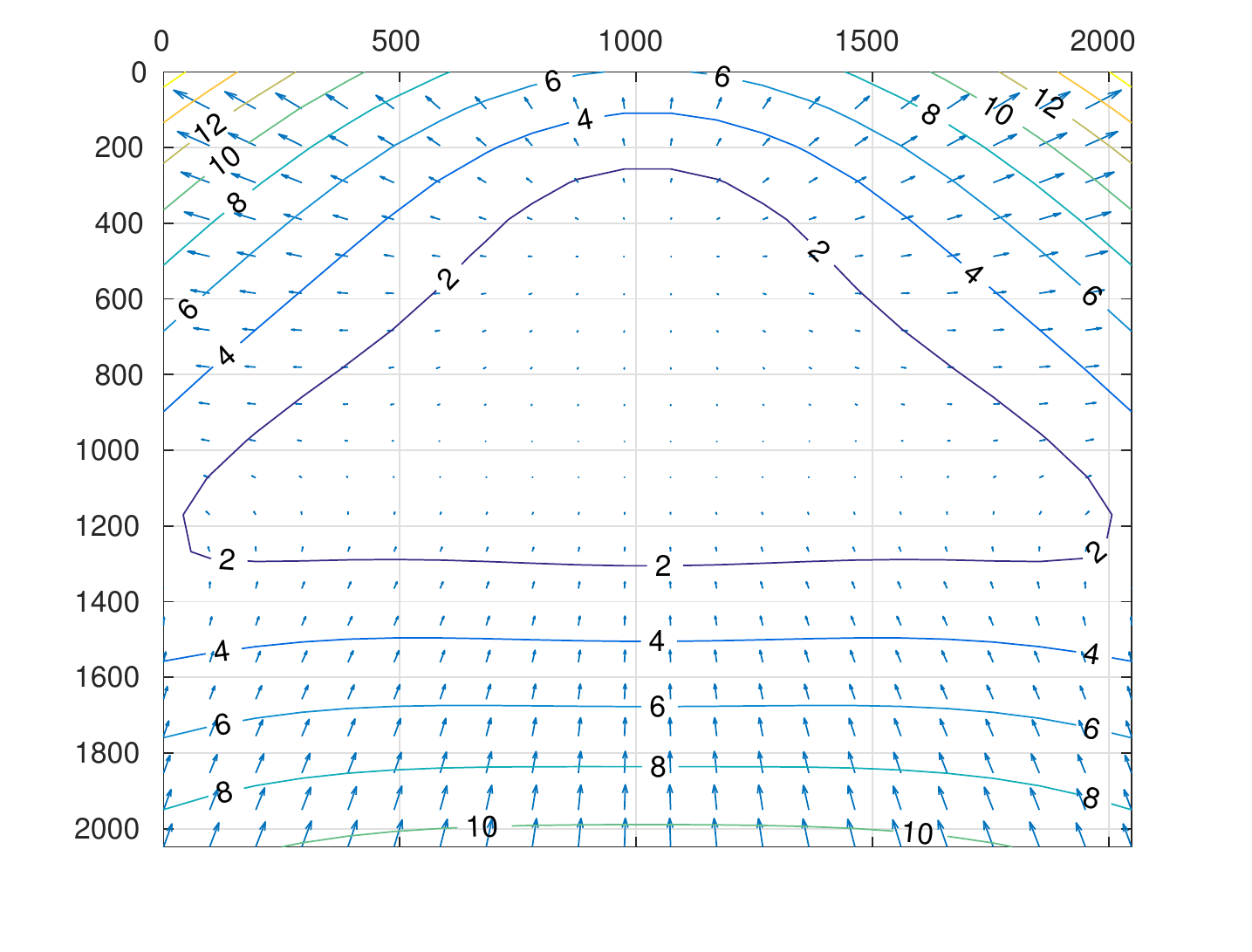}
  \caption{rational ( \(\overline{error}\)=0.088 pix)}
  \label{fig:simulated_rational_model}
\end{subfigure}%
\begin{subfigure}{.49\textwidth}
  \centering
  \includegraphics[width=1\textwidth]{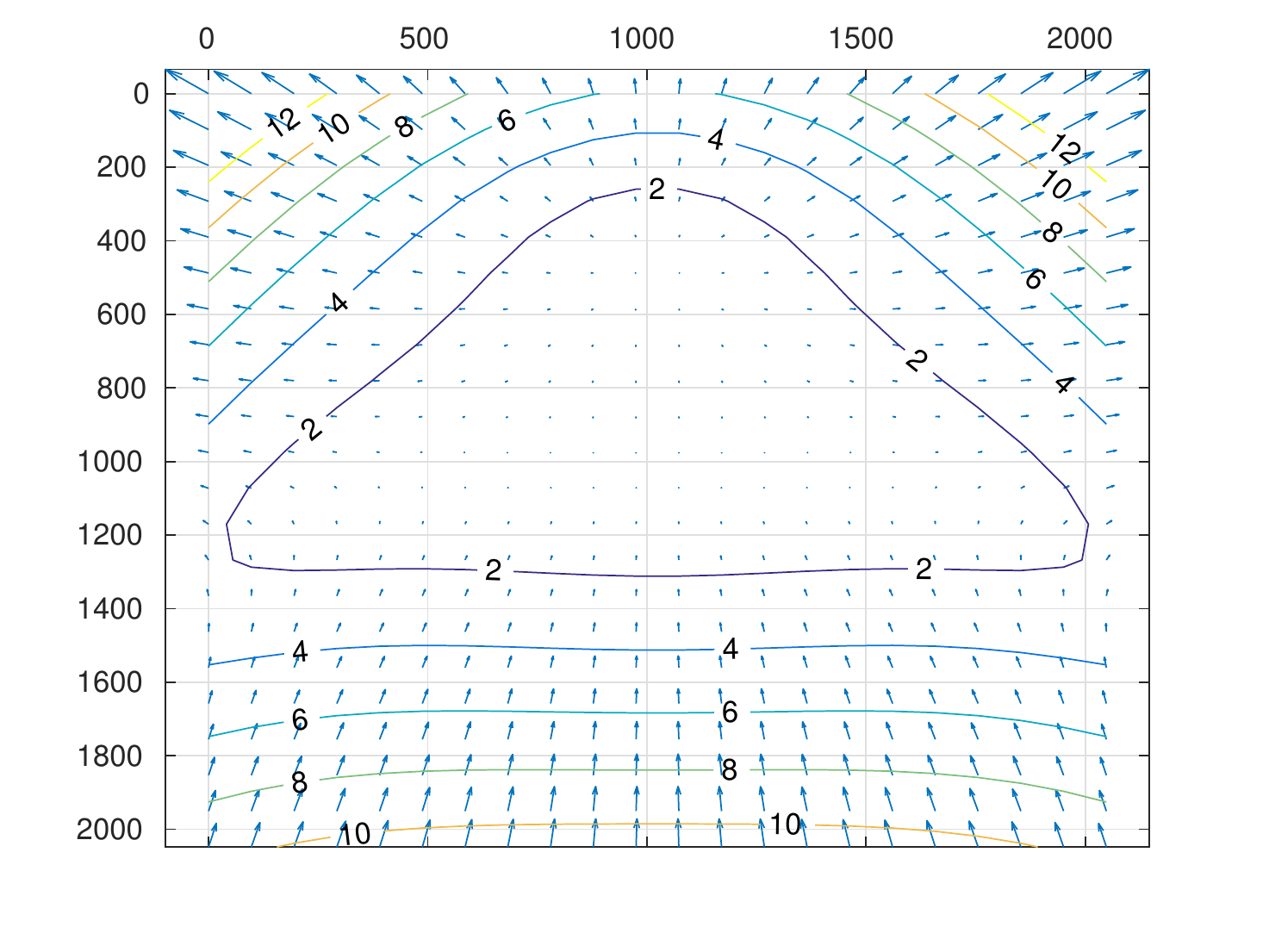}
  \caption{bi-cubic (\(\overline{error}\)=0.015 pix)}
  \label{fig:simulation_bicubic_model}
\end{subfigure}
\captionsetup{width=0.9\textwidth}
\caption{Distortion models ``fitted'' to simulated CaSSIS optical distortion data (see Tab.~\ref{tab:lens_simulation} in appendix). Vectors show the transformation from the distorted to the ideal image. Contour lines show the magnitude of this transformation. The errors are the average Euclidean distances between the positions of the ideal pixels, as predicted by the model, and their actual positions. Note that simple radial~(\ref{fig:simulation_radial_model}) and Brown-Conrady models~(\ref{fig:simulation_brown_conrandy_model}), with more than 1 pixel error, fail to represent CaSSIS distortion, while bi-cubic~(\ref{fig:simulation_bicubic_model}) and rational~(\ref{fig:simulated_rational_model}) models, with less than 0.1 pixels error, both perform well.
}
\label{fig:distortion_model_selection}
\end{figure}

The resulting distortion fields and errors for each model are shown in Fig.~\ref{fig:distortion_model_selection}. Simple radial and Brown-Conrady models suffer from more than 1 pixel error, and hence failed to represent the CaSSIS distortion, while bi-cubic and rational models, with less than 0.1 pixel error, performed well. We decided to use the rational model.

\section{On-ground calibration}
\label{sec:on_ground_calibration}

\begin{figure}[h!]
\begin{minipage}{.6\textwidth}
\centering
\includegraphics[width=1\textwidth]{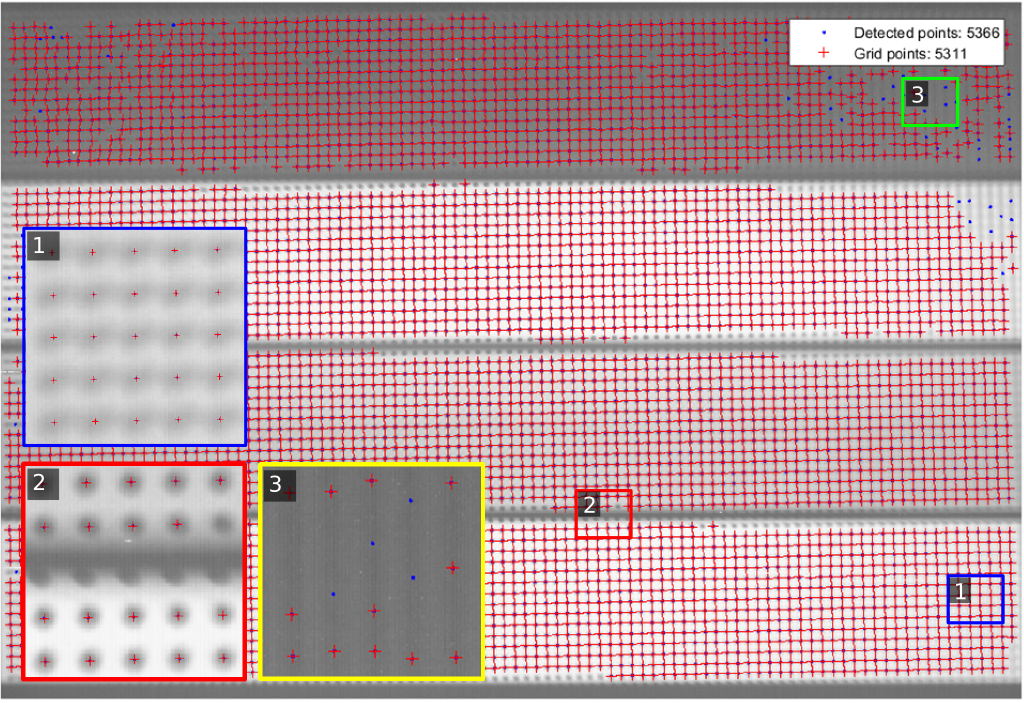}
  \caption{Image of the dotted target overlaid with the fitted grid. Red crosses show dots that were added to the grid. Blue points show dots, that were not added to the grid.}
 \label{fig:dots_detection_grid_fitting}
 \end{minipage}
 \begin{minipage}{.4\textwidth}
  \centering
  \captionsetup{width=0.9\textwidth}
  \includegraphics[width=0.95\textwidth]{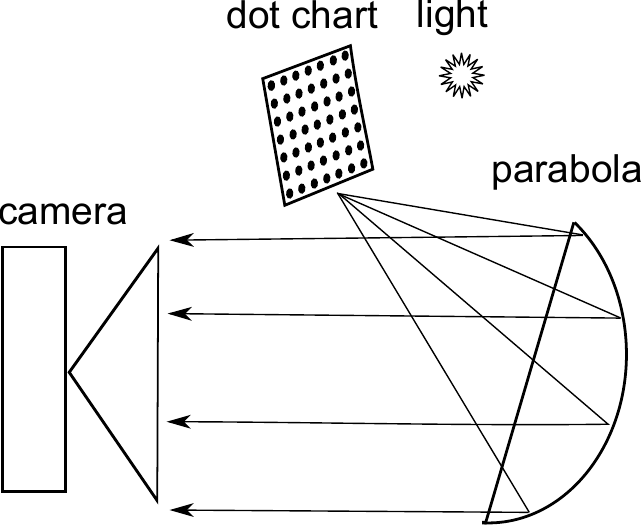}
  \captionof{figure}{On-ground calibration setting. To acquire in-focus image of the dotted calibration target from a reasonable distance, we put it in the focus of the parabolic collimator. }
 \label{fig:settings_with_collimator}
\end{minipage}%

\end{figure}

After the CaSSIS camera was assembled and tested, we attempted to estimate the distortion model from a single image of a dotted calibration target, as in~\citealt{claus2005}. Because the focal length of CaSSIS is too large to acquire in-focus images of the target from a reasonable distance, we used a set-up with a collimator (Fig.~\ref{fig:settings_with_collimator}).

After the image was acquired, we applied adaptive thresholding and connected components detection methods to identify dots in the image. Then, we found the dots' centers using a centroid algorithm. Finally, we fitted the regular rectangular grid to the dots' centers, using a simple algorithm that starts from an arbitrarily-selected dot, and expands the grid in horizontal and vertical directions, until no new dots can be added to the grid.

The acquired image with the fitted grid is shown in Fig.~\ref{fig:dots_detection_grid_fitting}. Analysis of the grid confirmed the presence of a small optical distortion in the image: the grid rows and columns appeared, not as straight lines but, as high-order curves. However, we failed to estimate the distortion field resembling Fig.~\ref{fig:simulated_rational_model} using the grid. This is probably due to the fact that the experimental data was contaminated with a then-unknown residual distortion coming from the off-axis collimator.

\section{In-flight calibration}
\label{sec:in_flight_calibration}

During TGO commissioning and mid-cruise checkout, CaSSIS acquired multiple images of star fields, that we used for in-flight calibration. In~\S~\ref{sec:inflight_calibration_method} we describe our in-flight calibration method and in~\S~\ref{sec:inflight_calibration_results} we show the calibration results.

\subsection{Method}
\label{sec:inflight_calibration_method}

\begin{figure}[htb]
\begin{minipage}{.5\textwidth}
  \centering
  \captionsetup{width=0.9\textwidth}
  \includegraphics[width=0.95\textwidth]{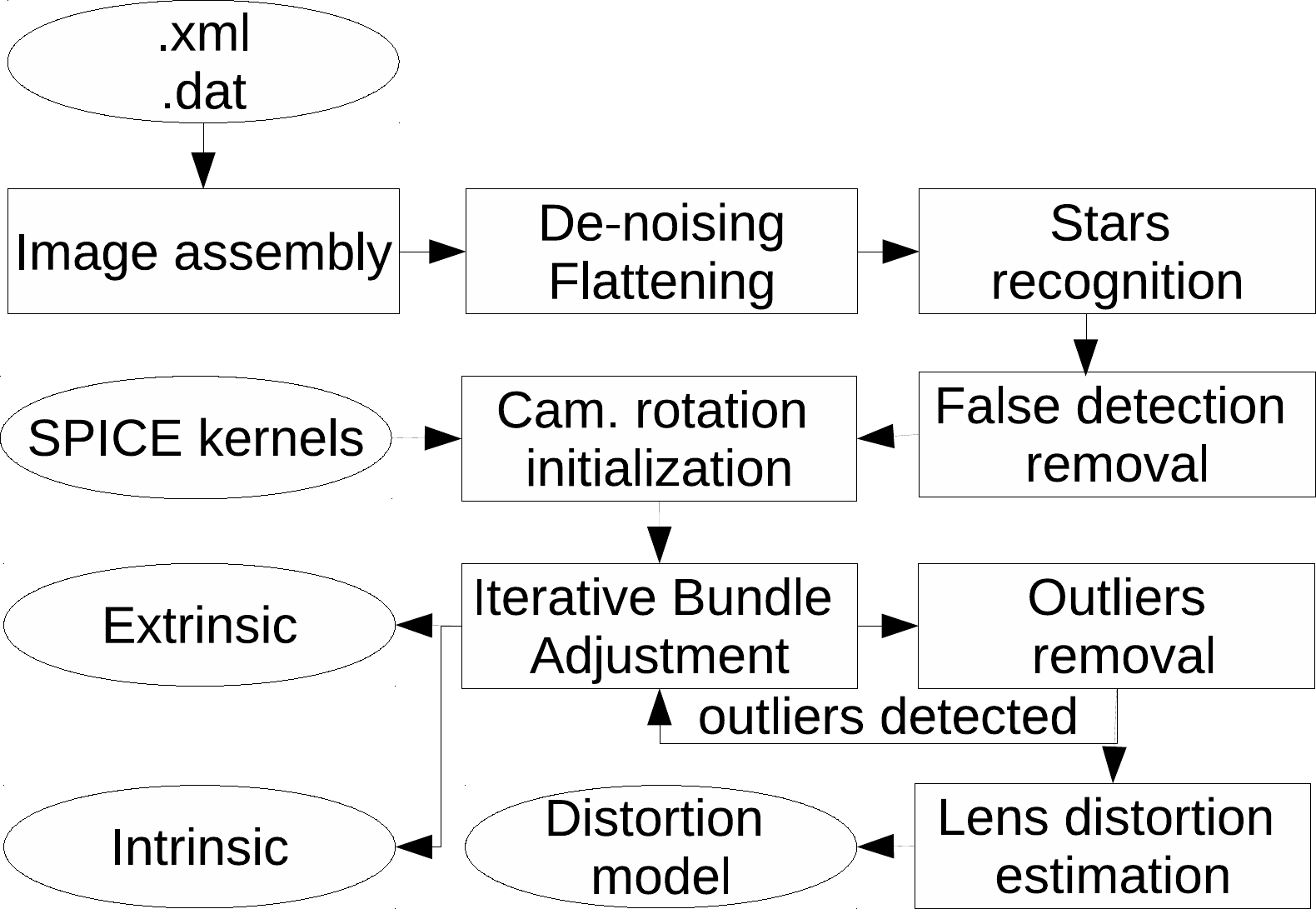}
  \captionof{figure}{Work flow of the in-flight calibration. Ellipses show the input and the output data and rectangles show processing steps.}
  \label{fig:starfield_calibration_workflow}
\end{minipage}
\begin{minipage}{.5\textwidth}
  \centering
  \captionsetup{width=0.98\textwidth}
  \includegraphics[trim={0.5cm 1cm 1cm 0.25cm},clip,width=0.99\textwidth]{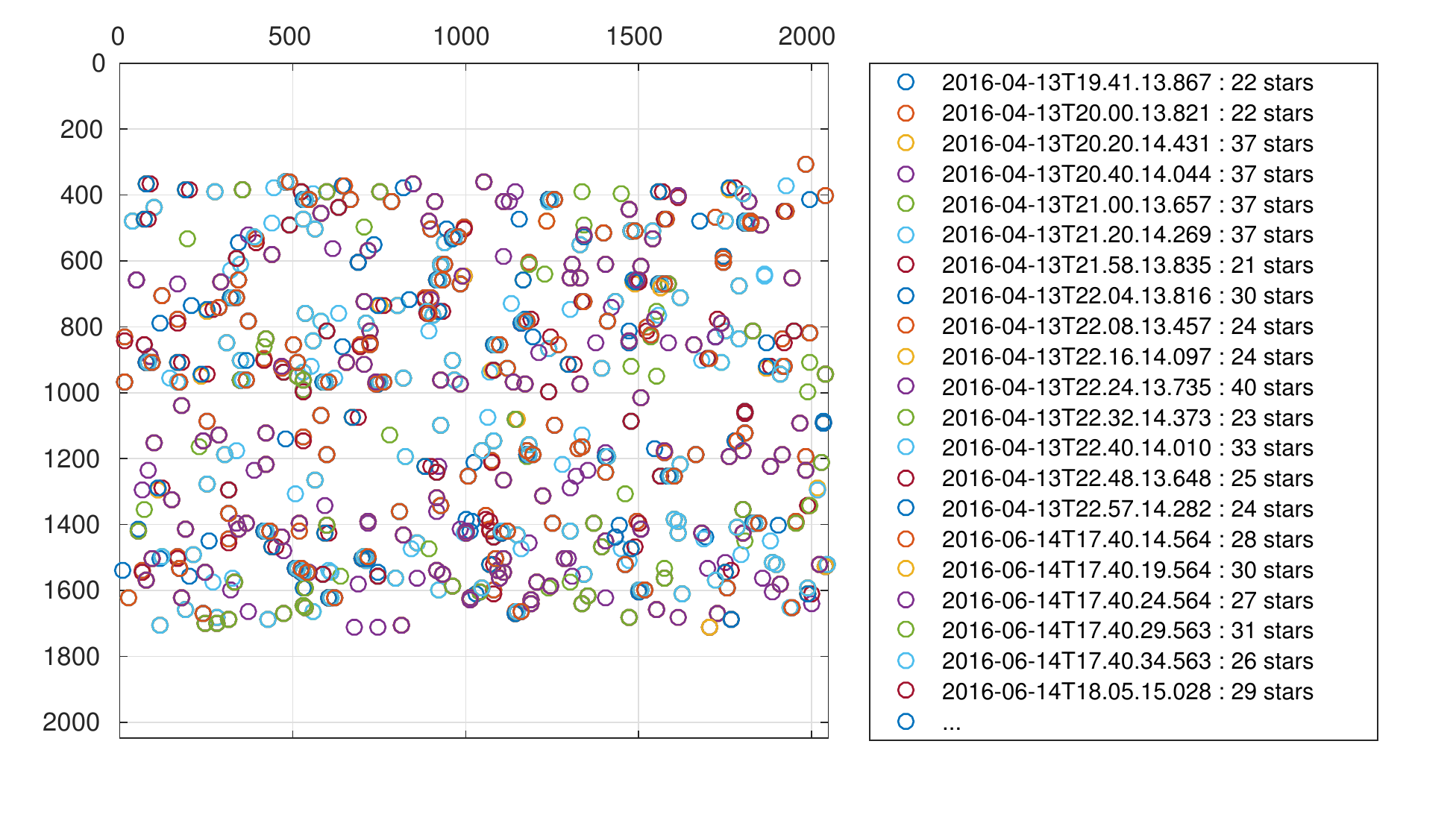}
   \captionof{figure}{Position of all stars detected in combined ``mcc motor'' + ``pointing CaSSIS'' set on the image sensor. Note that the detector is almost uniformly filled. On the top and the bottom parts of the sensor we don not have observations, since they are covered by nontransparent mask.  }
   \label{fig:starfield_density}
\end{minipage}%
\end{figure}

The overall work flow for in-flight calibration is shown in  Fig.~\ref{fig:starfield_calibration_workflow} and each individual procedure is described below. They are performed in that order.

\paragraph*{Image assembly.} We assemble full-sensor images from several data packets according to information in XML files from the telemetry conversion (Each image is accompanied by housekeeping data)

\paragraph*{De-noising and flattening.} We denoise every image by subtracting the median of several images from each image. This procedure helps us to get rid of fixed-pattern noise and hot pixels. Then we flatten each image by applying a Difference-of-Gaussian (DoG) filter.

\paragraph*{Star field recognition.} We perform star detection and recognition using the open-source Astrometry.net library and 2MASS star catalog. The library takes an image of a star field as an input, and outputs $ (x,y) $ coordinates of stars in the image, and their corresponding $ (Ra, Dec) $ coordinates in equatorial frame J2000.

\begin{wraptable}{r}{0.48\textwidth}
\centering 
\resizebox{0.49\columnwidth}{!}{%
\begin{tabular}{llcc}
\hline
Data & Name & \begin{tabular}{c} No.\\images \end{tabular} & \begin{tabular}{c} No.\\stars \end{tabular} \\
\hline
2016-04-13 & pointing cassis & 45 & 539 \\
2016-06-14 & mcc motor & 92 & 2573 \\
\rowcolor{gray}
2016-04-07 & commissioning 2 & 12 & 670 \\
\hline
\end{tabular}
}
\caption{Datasets summary. Note that the calibration sets consist of sequences of 3-4 almost identical images, acquired within short time interval. There are  10-60 stars in each image.}
\label{tab:dataset_summary}
\end{wraptable}

\paragraph*{False detections removal} In the next step we collect information about detected stars from all images and filter out erroneous detections. Since the calibration image sets consist of sequences of 3-4 almost identical images, we mark a star as a false detection, if it is not re-detected at a similar position in at least 2 images.

\paragraph*{Camera rotation initialization} We find the camera rotation for every image independently. During the estimation, we set the focal length of the camera to nominal and search for the camera rotation that minimizes the projection error, i.e the Euclidean distance between observed and predicted star positions in each image individually. The optimization is done with Levenberg–Marquardt algorithm~(lsqnonlin in MATLAB). We initialize the optimization with rotation angles from the SPICE kernel\footnote{ExoMars Trace Gas Orbiter SPICE kernels, \url{https://naif.jpl.nasa.gov/pub/naif/EXOMARS2016
}, accessed: 2017-05-24}\footnote{SPICE toolkit,
  \url{https://naif.jpl.nasa.gov/naif/toolkit.html
}, accessed: 2017-05-24}. The SPICE kernels contain information about orientation and position of spacecraft and its elements, received from its sensors for any moment in time.

\begin{wrapfigure}{r}{.48\textwidth}
 \centering
 \includegraphics[trim={0.5cm 1.2cm 0.5cm 0.3cm},clip,width=0.5\textwidth]{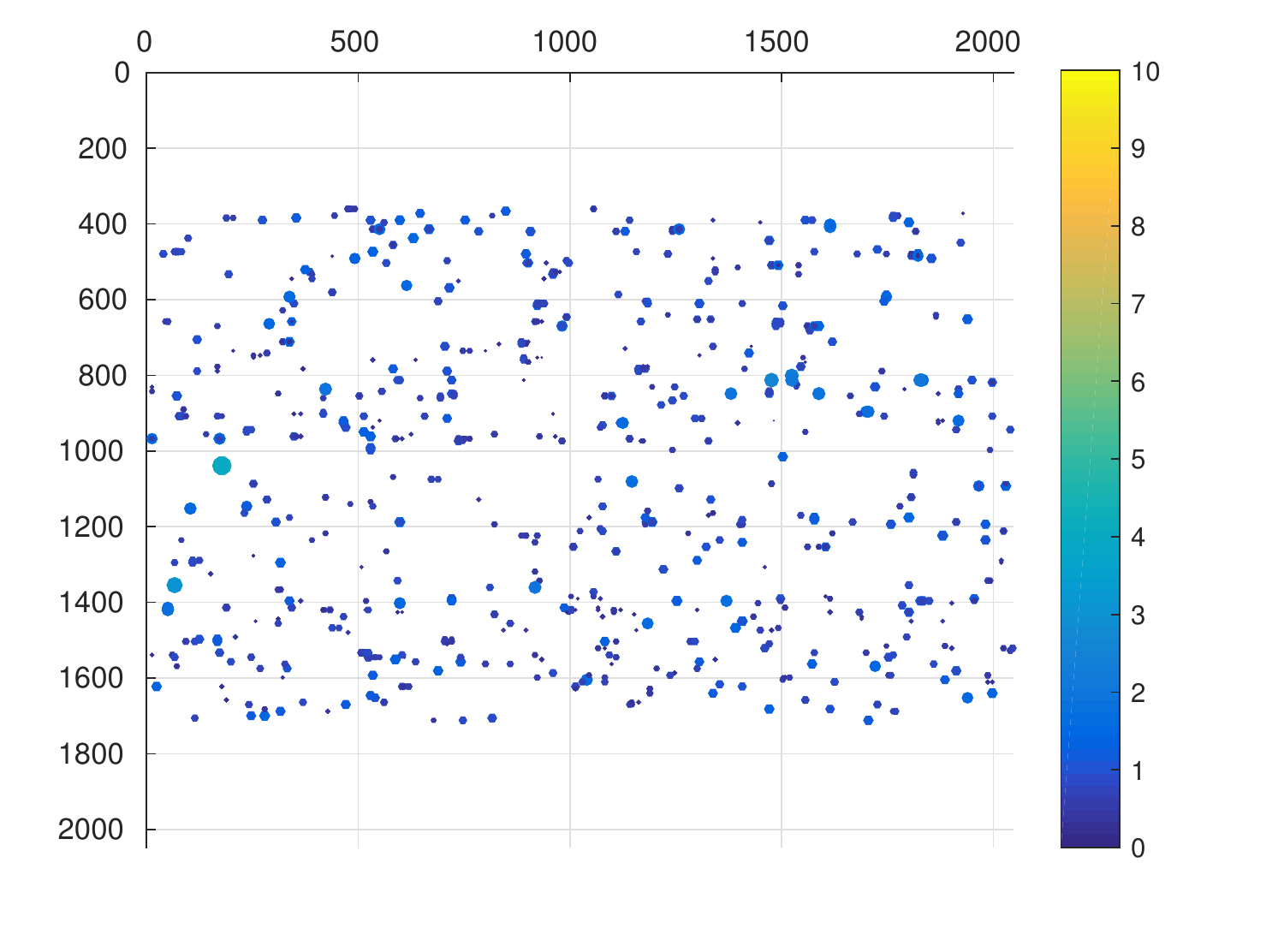}
\captionsetup{width=0.49\textwidth}
\captionof{figure}{Residual errors in pixels after optical distortion estimation. The average error is 0.66 pixels. Color coding shows the actual error scale, that is similar to Fig.~\ref{fig:ba_residuals_iteration4}. Note that the errors are small and spatially more uniform than compared to the residual errors after BA from Fig.~\ref{fig:ba_residuals_iteration4}. This suggests that they come from inaccurate star detection.}
\label{fig:distortion_model_residual}
\end{wrapfigure}%

\paragraph*{Iterative Bundle Adjustment~(BA)} In this step we search for a refined focal length and rotations that minimize the projection errors for all images simultaneously. The optimization is performed with Levenberg–Marquardt algorithm. We initialize the optimization with the focal length and rotation matrices we found in the previous step. After each BA iteration, stars that have large residual projection errors compared to their spatial neighbors are rejected as outliers and BA is performed again until no new outliers are found. Without this outlier rejection, the subsequent optical distortion estimation would fail.

\paragraph*{Rational optical distortion estimation} In this step we ``freeze'' the intrinsic and the extrinsic camera models and search for a rational optical distortion model that minimizes the remaining projection error. The optimization is performed with Levenberg–Marquardt algorithm. We initialize the optimization process using a ``no distortion'' hypothesis.

\subsection{Results}
\label{sec:inflight_calibration_results}

We performed our experiments on 3 datasets: ``mcc motor'' and ``pointing cassis'' both acquired in July 2016 during mid-cruise checkout, and ``commissioning2'', acquired on April 2016 during near-Earth commissioning. We selected these datasets, since they contain images of dense star fields acquired with long 1.92-second exposures. We estimated the camera parameters using the combined ``mcc motor'' and ``pointing cassis'' set, which we called the {\it training set}, and validated the results on ``commissioning2'' set, which we called the {\it validation set.}

\begin{figure}[htp!]
\centering
\begin{subfigure}{.49\textwidth}
 \includegraphics[trim={0.0cm 1.2cm 2.7cm 0.3cm},clip,width=0.99\textwidth]{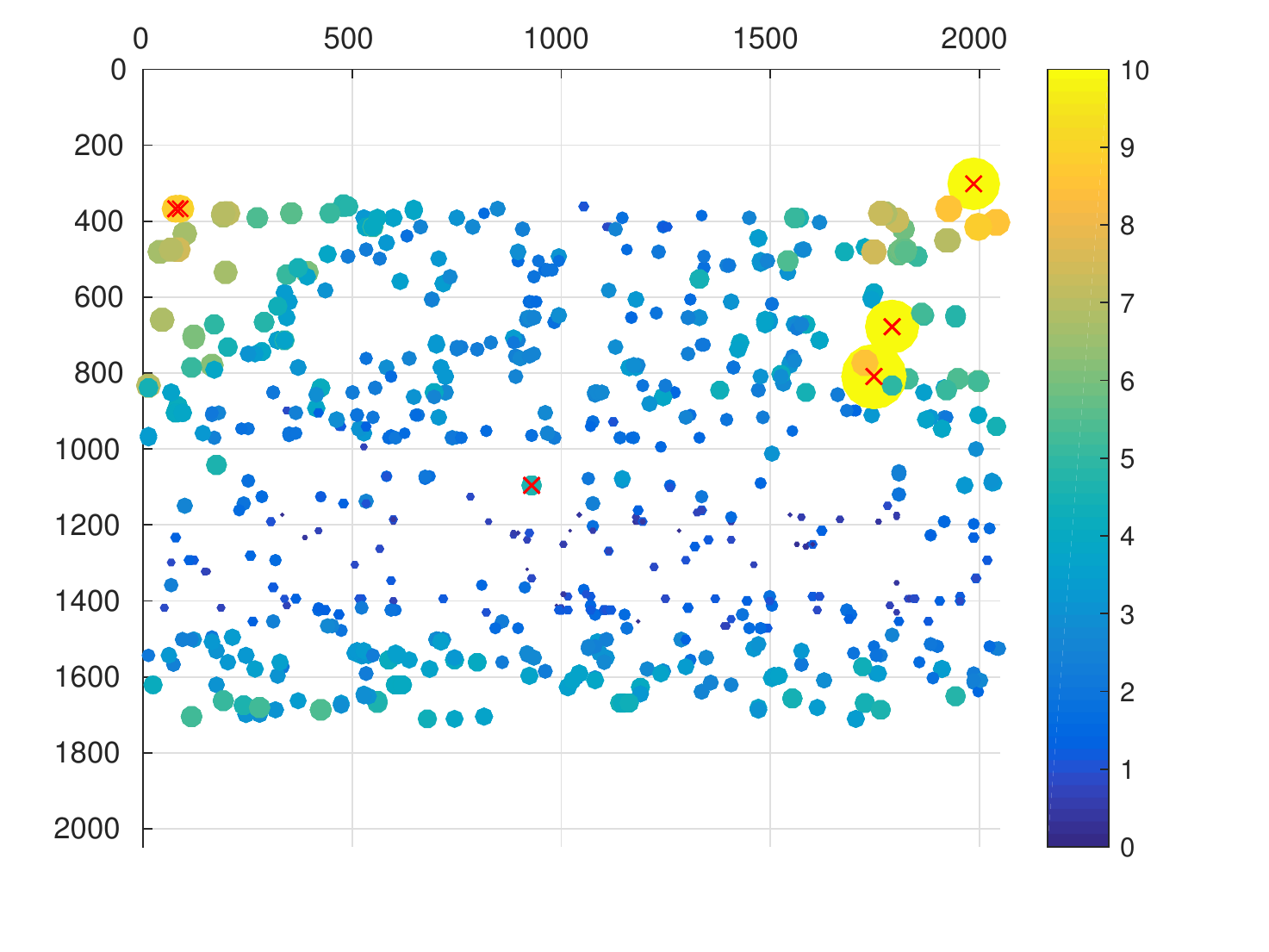}
\caption{$1^{st} $ BA iteration}
\label{fig:ba_residuals_iteration1}
\end{subfigure}%
\begin{subfigure}{.49\textwidth}
  \centering
 \includegraphics[trim={0.9cm 1.2cm 1.3cm 0.3cm},clip,width=0.99\textwidth]{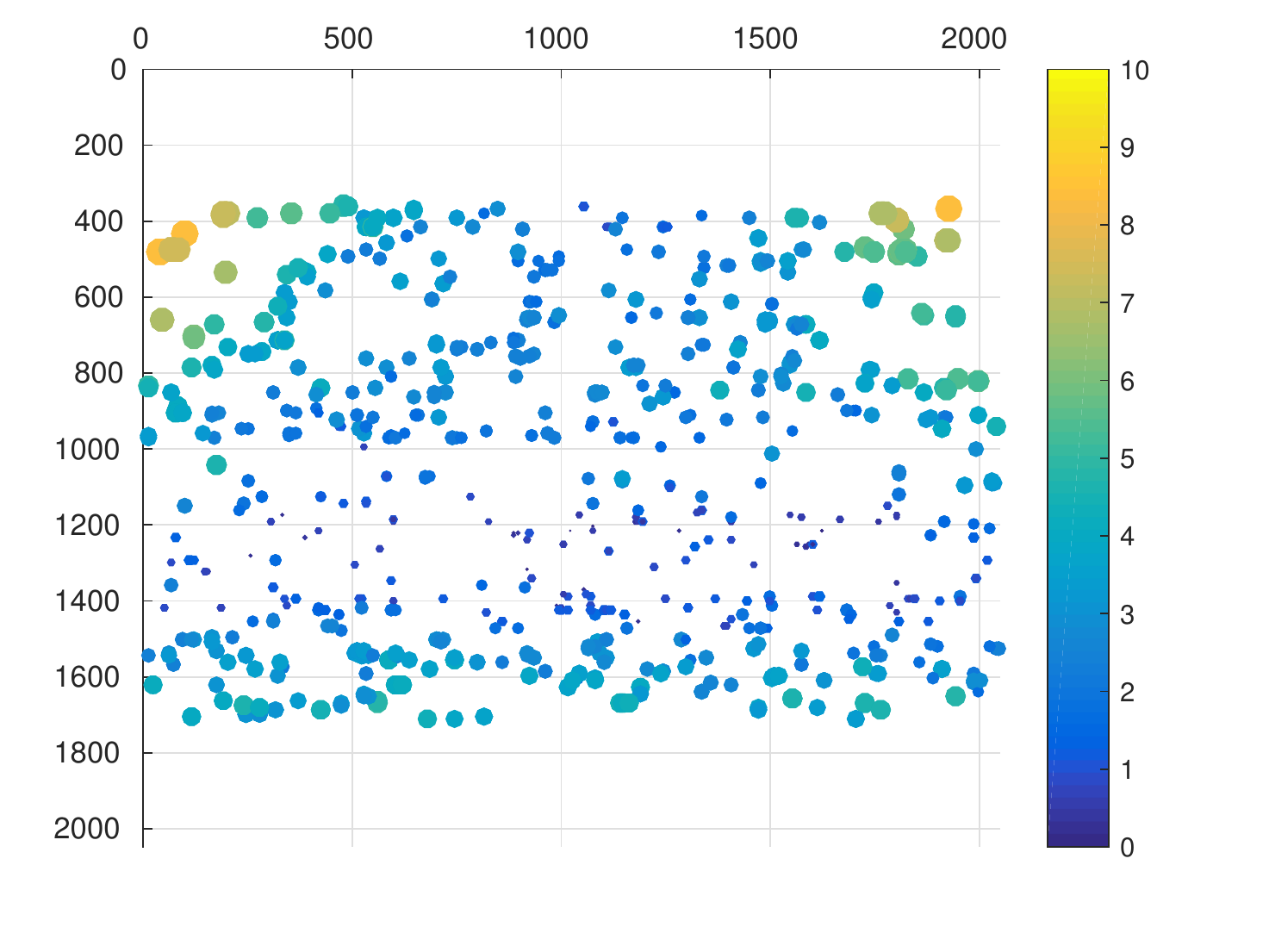}
\caption{$4^{th}$ (final) BA iteration }
\label{fig:ba_residuals_iteration4}
\end{subfigure}%
\captionsetup{width=0.95\textwidth}
\caption{Residuals after the first and the forth BA iterations. Color coding shows the actual scale of the residuals. Crossed-out residuals correspond to the identified outliers. On the top and the bottom parts of the sensor we do not have observations, since they are covered by a nontransparent mask. Note that after the first iteration~(\ref{fig:ba_residuals_iteration1}), the residuals contain gross outliers, while after the forth iteration~(\ref{fig:ba_residuals_iteration4}) residuals form a clear spatial pattern suggesting the presence of optical distortion. The average Euclidean error before BA is 3.65 pixels, after the first iteration it is 2.78 pixels, and after the forth iteration it is 2.56 pixels.}
\label{fig:effect_of_iterations}
\end{figure}

A number of images and the recognized stars in every set are shown in Tab.~\ref{tab:dataset_summary}.
As shown in Fig.~\ref{fig:starfield_density}, the stars from the training set cover the sensor densely and uniformly, allowing for good optical distortion estimation.

Using stars detected in the training set, we refined the camera rotations obtained from SPICE kernels for every image individually, while keeping the focal length of the camera fixed to the nominal. By refining the rotations, we reduced the median Euclidean distance between observed and the predicted star positions in training-set images from 147.41 to 3.42 pixels.

Then, we used the estimated camera rotations and the nominal focal length to initialize the iterative bundle adjustment process that refined the camera rotations and focal length using all images simultaneously, while ignoring optical distortion. The iterative bundle adjustment converged after 4 iterations. The effect of the iterative outliers rejection scheme is shown in Fig.~\ref{fig:effect_of_iterations}. Note that after the first iteration, the BA residuals contain gross outliers, while after the last iteration, the residuals form a clear spatial pattern suggesting the presence of optical distortion. BA reduced the average Euclidean distance between observed and predicted star positions in the training set images from 3.56 to 2.56 pixels. The refined focal length was found by BA to be 875.93 mm, i.e. slightly shorter than the nominal focal length of 880 mm.

\begin{wraptable}{r}{0.5\textwidth}
\centering
\resizebox{0.5\textwidth}{!}{%
\begin{tabular}{llllll}
\rowcolor{gray}
A11 & A12 & A13 & A14 & A15 & A16 \\
0.0643 &   0.4091 &   -0.0011 &   1.0003  &  0.0003 &  -0.0000 \\
\rowcolor{gray}
A21 & A22 & A23 & A24 & A25 & A26 \\
-0.0043 & 0.0635  & 0.4065  & 0.0002 & 0.9952 & 0.0004 \\
\rowcolor{gray}
A31 & A32 & A33 & A34 & A35 & A36 \\
-0.0501 & 0.0071 & -0.0305 & 0.0636 & 0.4401 & 1.0000 \\
\end{tabular}
}
\caption{Parameters of the rational distortion model from \S~\ref{sec:rational_model}, estimated using star field images.}
\label{tab:distortion_model_parameters}
\end{wraptable}

\begin{figure}[hbtp]
\centering
\begin{subfigure}{.49\textwidth}
\includegraphics[trim={0.5cm 1.2cm 0.5cm 0.3cm},clip,width=0.99\textwidth]{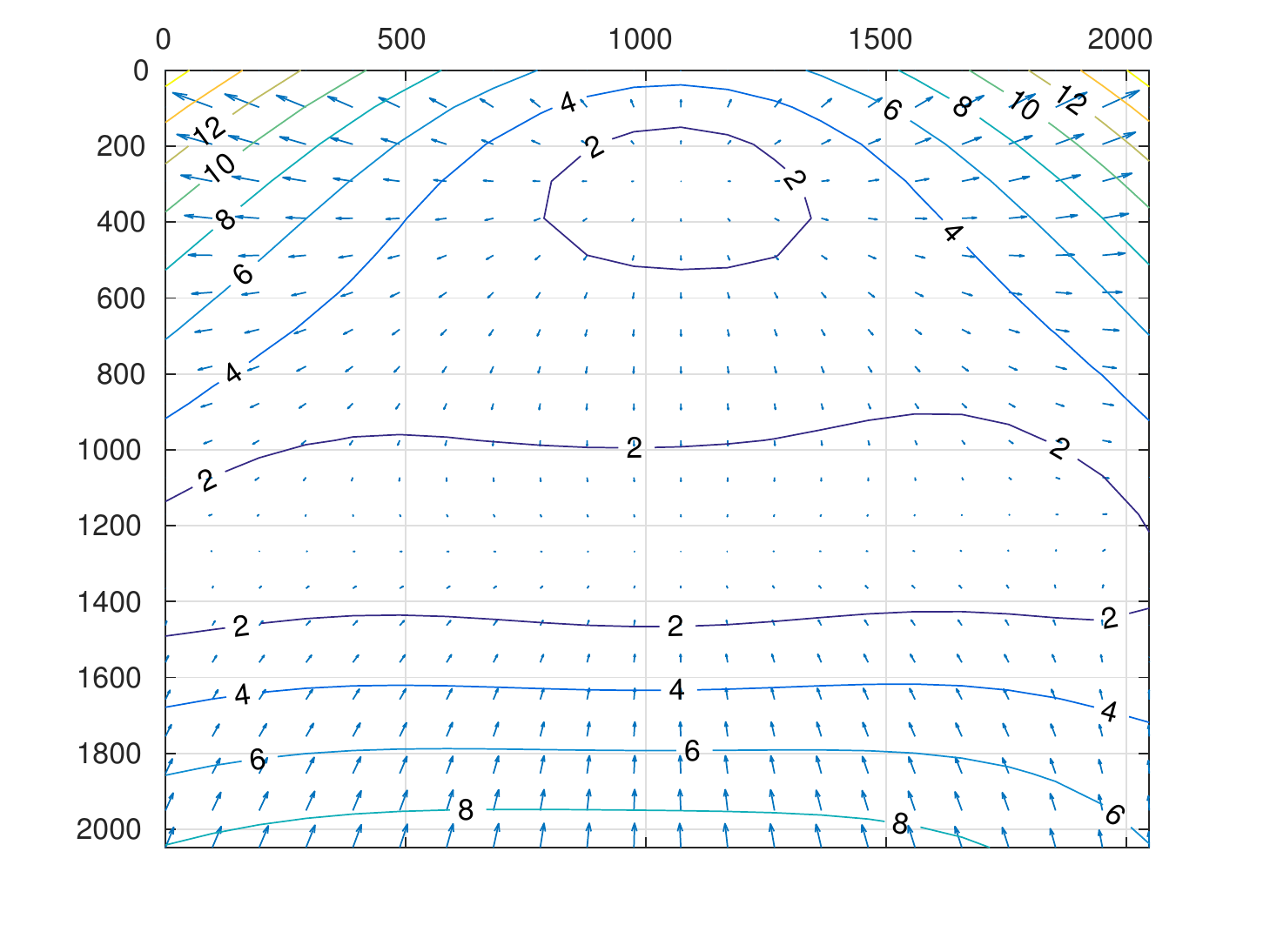}
\caption{from star field images}
\label{fig:distortion_from_star_field}
\end{subfigure}%
\begin{subfigure}{.49\textwidth}
\includegraphics[trim={0.5cm 1.2cm 0.5cm 0.3cm},clip,width=0.99\textwidth]{simulation_rational_model.pdf}
\caption{from simulation}
\label{fig:distortion_from_simulation}
\end{subfigure}%
\captionsetup{width=0.95\textwidth}
\caption{The distortion field estimated from star field images~(\ref{fig:distortion_from_star_field}) and optical simulations~(\ref{fig:distortion_from_simulation}), described in~\S\ref{sec:distortion_model_selection}. Vectors show transformation from distorted to ideal image. Contours show magnitude of the transformation. Note that the distortion fields are very similar in shape, with an apparent vertical translation of the field as the most obvious difference..}
\label{fig:distortion_fields_comparision}
\end{figure}

Then, we ``froze'' the focal length and camera rotations and estimated the rational distortion model. The estimated distortion field is shown in Fig.~\ref{fig:distortion_from_star_field}. Note that its shape resembles the distortion field obtained by fitting the rational model to lens simulation data in~\S\ref{sec:distortion_model_selection}, duplicated for convenience in Fig.~\ref{fig:distortion_from_simulation}.

Parameters of the estimated distortion model are shown in Tab.~\ref{tab:distortion_model_parameters}. The distortion model fitting reduced the average Euclidean distance between the observed and the predicted star positions in the training set images from 2.54 to 0.66 pixels. Moreover, as shown in Fig.~\ref{fig:distortion_model_residual}, the residuals after fitting the optical distortion model became spatially uniform and small when compared to the bundle adjustment residuals from Fig.~\ref{fig:ba_residuals_iteration4}. This suggests that the remaining residual errors probably come from inaccurate star detection.

Finally, we computed the error of the estimated camera model on a separate validation set, that was not used for model estimation, while effectively ignoring extrinsic model. With the refined camera model, the average projection error is 0.47 pixels, while with the nominal camera model the error would be 3.56 pixels. This result suggests that our geometric calibration results are valid.

\section{Colour image experiment}
\label{sec:color_image_experiment}

\begin{wrapfigure}{r}{0.5\textwidth}
  \centering
  \captionsetup{width=0.5\textwidth}
  \includegraphics[width=0.49\textwidth]{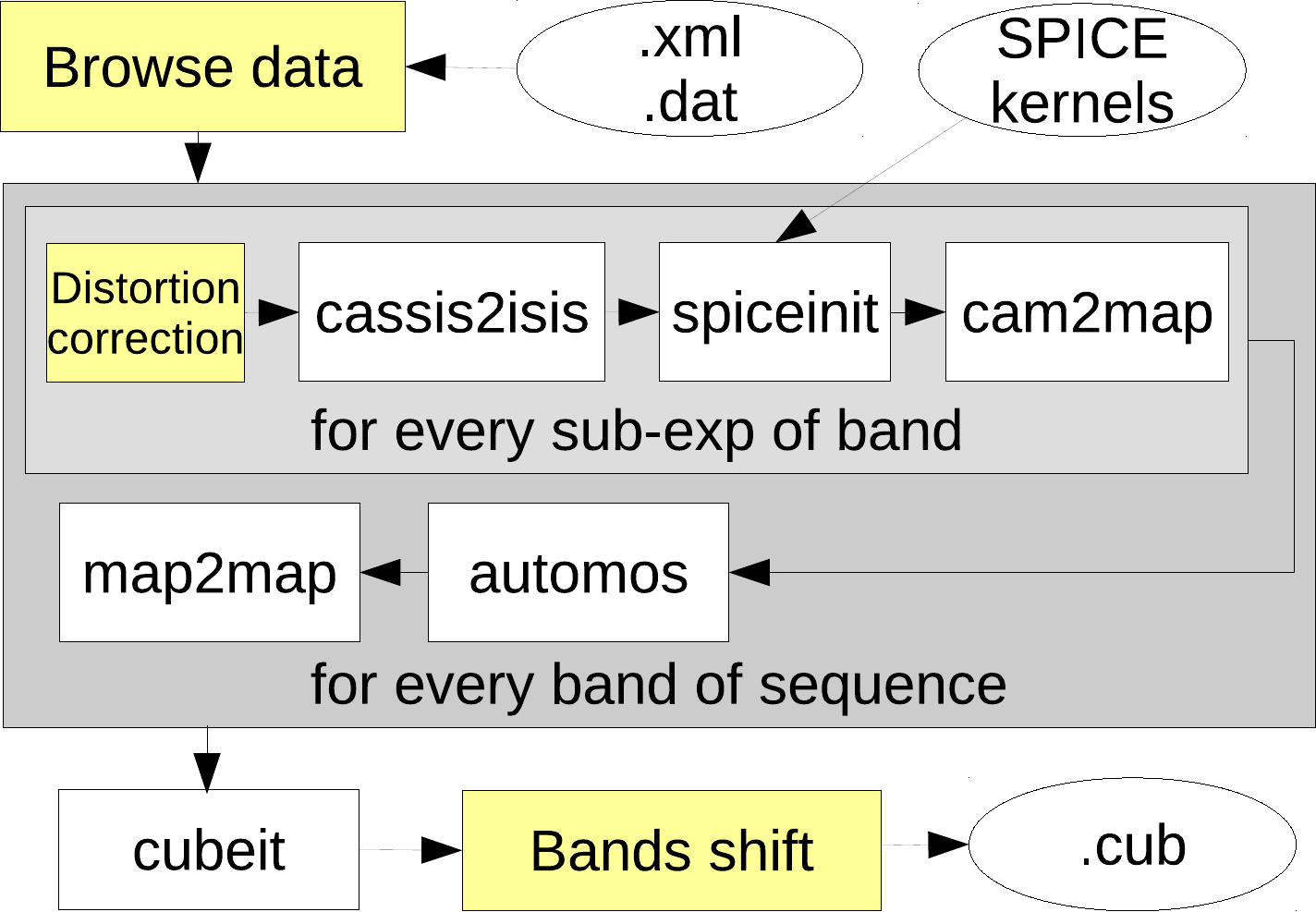}
  \captionof{figure}{Work flow of the color image experiment. Ellipses represent data, white rectangular boxes represent the standard ISIS functions, and yellow boxes represent the scripts implemented in Python.}
 \label{fig:workflow_color_image_experiment}
\end{wrapfigure}

A month after TGO's Mars orbital insertion, CaSSIS captured several colour images of Mars from the elliptic capture orbit. In order to verify the effectiveness of our calibration, we map-projected these images using the nominal and the refined geometric models, and compared the quality of the resulting images. The map-projection was performed using Integrated Software for Imagers and Spectrometers~(ISIS)\footnote{USGS Integrated Software for Imagers and Spectrometers, \url{https://isis.astrogeology.usgs.gov}, accessed: 2017-06-06}. In~\S~\ref{sec:color_image_experiment_method} we describe the work flow of our color image experiment, and in~\S~\ref{sec:color_image_experiment_results} discuss the results.

\subsection{Method}
\label{sec:color_image_experiment_method}

The work flow of the colour image experiment is shown in  Fig.~\ref{fig:workflow_color_image_experiment}, with each individual procedure described below.

First of all, all data packets belonging to a particular sequence and a color band are extracted from the dataset. Next, we correct the optical distortion in every data packet. Then, we convert each data packet to ISIS ``.cub'' format (\textit{cassis2isis}) and add information from the SPICE kernel to each ``.cub'' (\textit{spiceinit}). After that, we project all ``.cub'' that correspond to a single band of a image sequence into a sinusoidal map (\textit{cam2map}), while keeping the resolution of the projections consistent. Next, we mosaic all projected ``.cub'' into one image, corresponding to a single band of the sequence (\textit{automos}).
We repeat the process described above for every color band. After that, we select one of the bands as a reference and match map-projections of all other bands to it (\textit{map2map}). This is required since ``by default'' map projection of every band has its own resolution and coordinate limits. Finally, we combine the individual color bands into a multi-band cube (\textit{cubeit}).

\subsection{Results}
\label{sec:color_image_experiment_results}

During our experiments we noticed that map-projected images of individual color bands were misaligned along the track by 1-10 pixels, depending on the sequence. This fact, possibly caused by the off-nadir pointing of the camera relative to its rotation axis being slightly different from the nominal. This fact needs to be properly investigated. Meanwhile we checked the validity of the optical distortion model by verifying that the color band images are distortion-free. For that we compensated color band misalignment with a simple shift and compared color band images.

\begin{figure}[hbtp]
\centering
\begin{subfigure}{0.9\textwidth}
  \centering
  \includegraphics[trim={0.25cm 1.3cm 0.5cm 0cm},clip, width=1\textwidth]{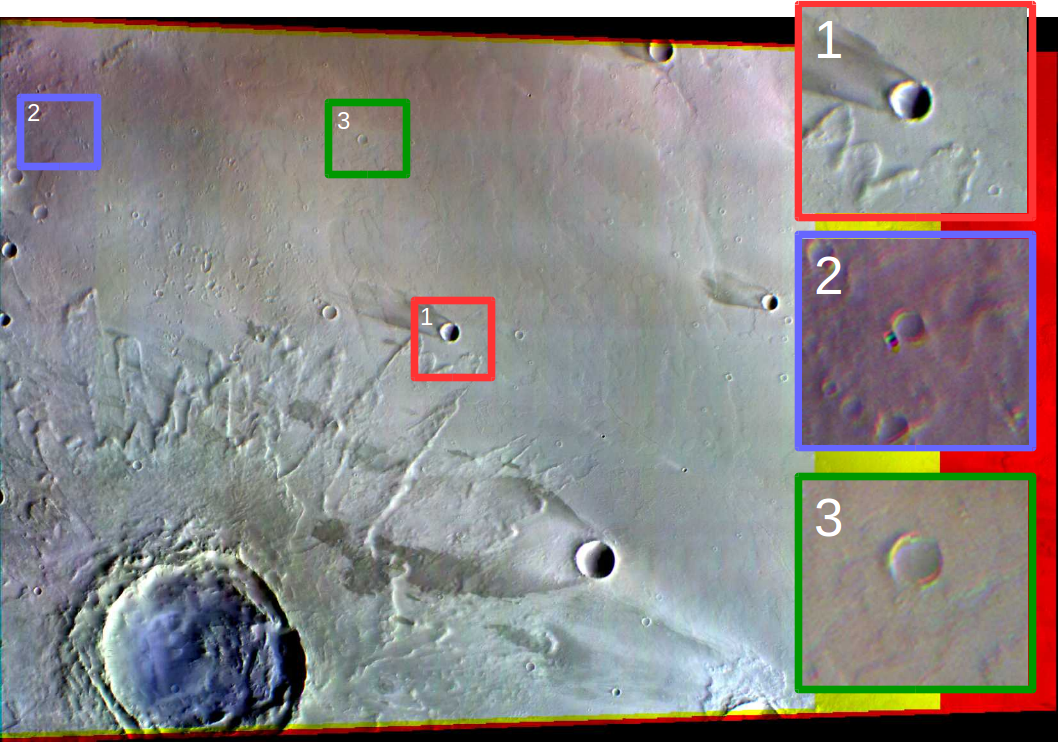}
  \caption{Nominal parameters}
  \label{fig:color_image_experiment_nominal}
\end{subfigure}
\begin{subfigure}{0.9\textwidth}
  \centering
  \includegraphics[trim={0.25cm 1.2cm 0.5cm 0cm},clip,width=1\textwidth]{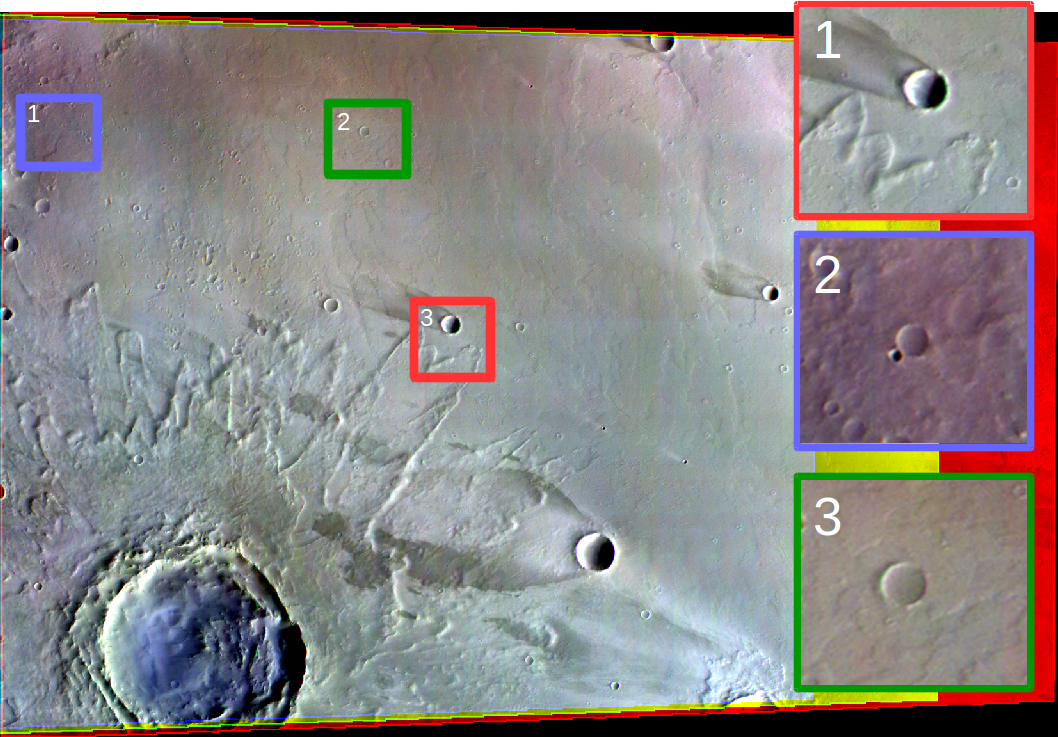}
  \caption{Refined parameters}
  \label{fig:color_image_experiment_refined}
\end{subfigure}
\captionsetup{width=0.95\textwidth}
\caption{Map projection of a CaSSIS image (first ``framelet'' acquired on November 22, 2016 at 16:01:10, central $(lat, lon)=(-9.03, 218.49)$). Images are shown in false colour (RED $\rightarrow$ red, NIR $\rightarrow$ green, BLU $\rightarrow$ blue channel). On-ground resolution is 35.52 meters/pixel. Color bands are aligned as described in~\S\ref{sec:color_image_experiment_results}. Note that when we use nominal camera parameters, the projected image~(\ref{fig:color_image_experiment_nominal}) has color fringes (close-up \#1, 2 and 3) and stitching artifacts (close-up \#3), while when we use refined parameters, the projected image~(\ref{fig:color_image_experiment_refined}) is almost perfect. Some prominent artifacts in the form of vertical and horizontal banding over the image are due to incorrect photometric calibration, and not to incorrect geometric calibration. This issue is investigated independently.
}
\label{fig:color_image_experiment_results}
\end{figure}

The results of this comparison are shown in Fig.~\ref{fig:color_image_experiment_results}. As seen from the figure, when we use nominal camera parameters, the projected image~(\ref{fig:color_image_experiment_nominal}) has color fringes (close-up \#1, 2 and 3) and stitching artifacts (close-up \#3), whereas when we use refined parameters, the projected image~(\ref{fig:color_image_experiment_refined}) is almost perfect.

This confirms that the developed calibration method works and improves the quality of the final scientific products.

\section{Conclusion}

In this paper we developed a method for geometric calibration of telescopes with large focal length and complex optical distortion. The proposed method was used to refine the nominal parameters of the CaSSIS camera on board ESA's TGO. As a result, we were able to improve the quality of scientific products, such as color images.

Our method is general and can be used for the calibration of other telescopes. We further encourage re-use of the proposed method by making our calibration code and data available on-line.

\section*{Acknowledgements}
The authors wish to thank the spacecraft and instrument engineering teams for the successful completion of the instrument. CaSSIS is a project of the University of Bern and funded through the Swiss Space Office via ESA's PRODEX programme. The instrument hardware development was also supported by the Italian Space Agency (ASI) (ASI-INAF agreement no.I/018/12/0), INAF/Astronomical Observatory of Padova, and the Space Research Center (CBK) in Warsaw. Support from SGF (Budapest), the University of Arizona (Lunar and Planetary Lab.) and NASA are also gratefully acknowledged. We also acknowledge support from the NCCR PlanetS.

\section*{Appendix}

\begin{table}[htbp]
\centering
\scriptsize
\begin{tabular}{|l|m{0.13\textwidth}| m{0.13\textwidth}|m{0.13\textwidth}|m{0.13\textwidth}|}
\hline
& \bfseries \(x\), [mm] & \bfseries \(i\), [mm] & \bfseries \(y\), [mm] & \bfseries \(j\), [mm]
    \begin{filecontents*}{simulation_optical_distortion_predict.csv}
realX, realY, idealX, idealY, errorX, errorY
0,0,0,0,0,0
0,-3.3911,0,-3.3846,0,-0.65
0,-6.7437,0,-6.7538,0,1.01
0,3.4094,0,3.3846,0,2.48
0,6.8165,0,6.7538,0,6.27
-5.1358,0.0022,-5.1385,0,0.27,0.22
-5.1207,-3.3866,-5.1385,-3.3846,1.78,-0.2
-5.1029,-6.737,-5.1385,-6.7538,3.56,1.68
-5.1478,3.4093,-5.1385,3.3846,-0.93,2.47
-5.1568,6.8142,-5.1385,6.7538,-1.83,6.04
-10.2482,0.0089,-10.2769,0,2.87,0.89
-10.2183,-3.3733,-10.2769,-3.3846,5.86,1.13
-10.2133,-6.7171,-10.3077,-6.7538,9.44,3.67
-10.2722,3.4094,-10.2769,3.3846,0.47,2.48
-10.2901,6.8075,-10.2769,6.7538,-1.32,5.37
5.1358,0.0022,5.1385,0,-0.27,0.22
5.1207,-3.3866,5.1385,-3.3846,-1.78,-0.2
5.1029,-6.737,5.1385,-6.7538,-3.56,1.68
5.1478,3.4093,5.1385,3.3846,0.93,2.47
5.1568,6.8142,5.1385,6.7538,1.83,6.04
10.2482,0.0089,10.2769,0,-2.87,0.89
10.2183,-3.3733,10.2769,-3.3846,-5.86,1.13
10.183,-6.7173,10.2769,-6.7538,-9.39,3.65
10.2722,3.4094,10.2769,3.3846,-0.47,2.48
10.2901,6.8075,10.2769,6.7538,1.32,5.37
\end{filecontents*}
\csvreader[head to column names]{simulation_optical_distortion_predict.csv}{}
{\\\hline\thecsvrow&\csvcoli&\csvcoliii&\csvcolii&\csvcoliv}
\\\hline
\end{tabular}
\caption{The CaSSIS optical distortion data, computed using a ray-tracing simulation. $x$, $y$ are ideal-image coordinates and $i$, $j$ are distorted-image coordinates given relative to the image center. } \label{tab:lens_simulation}
\end{table}

\clearpage

\section*{References}


\checknbdrafts

\end{document}